\documentclass[twoside,a4paper,11pt,reqno]{amsart}
\usepackage{amsmath}
\usepackage{epsfig} \usepackage{psfrag}\usepackage{subfigure} 
\usepackage{tabularx}\usepackage{rotating}
\numberwithin{equation}{section}

\newtheorem{theorem}{Theorem}[section]
\newtheorem{lemma}[theorem]{Lemma}
\theoremstyle{definition}

\theoremstyle{remark}

\numberwithin{equation}{section}

\begin{document}

\title[Lyapunov Functionals for the Enskog Equation]
{Lyapunov Functionals for the Enskog Equation}

\author{Zhenglu Jiang}
\address{Department of Mathematics, Zhongshan University, 
Guangzhou 510275, P.~R.~China}
\email{mcsjzl@mail.sysu.edu.cn}
\thanks{ZJ is supported by NSFC 10271121, 10511120278 and 10611120371,  and sponsored by 
SRF for ROCS, SEM}



\date{August 27, 2006.}



\begin{abstract} 
Two  Lyapunov functionals are presented for the Enskog equation. 
One is to describe interactions between particles with various velocities 
and another is to measure the $L^1$ distance between two classical solutions. 
The former yields the time-asymptotic convergence 
of global classical solutions to the collision free motion 
while the latter is applied into the verification 
of the $L^1$ stability of global classical solutions.
\end{abstract}

\maketitle
\vspace*{-0.7cm}
\section{Introduction}
\label{intro}
We are concerned with two Lyapunov functionals to 
describe the time-asymptotic behaviour and 
the $L^1$ stability of global classical solutions to 
the Enskog equation without any external force 
for a hard sphere gas.   
In the absence of any external force, the Enskog equation is as follows: 
\begin{equation}
\frac {\partial f}{\partial t}+v\frac{\partial f}{\partial x}=Q(f)
\label{ee}
\end{equation}
where $ {f=f(t, x,v)}$ is a one-particle distribution function that denpends on 
the time $t\in {\bf R}_+,$ the position $x\in{\bf R}^3$ and the velocity   
$v\in{\bf R}^3,$  and 
$Q$ is the Enskog collision operator whose form will be addressed below. 
 Here and throughout this paper, 
 ${\bf R}_+$ represents the positive side of the real axis including its origin and 
 ${\bf R}^3$ denotes a three-dimensional Euclidean space.

The collision operator $Q$ is expressed by the difference between the gain and loss terms 
respectively defined by 
 \begin{equation}
Q^+(f)(t,x,v)=a^2\int_{{\bf R}^3\times S_+^2}f(t,x,v^\prime)f(t,x-a\omega,w^\prime)B(v-w, \omega) d\omega dw,
\label{eekp}
\end{equation}
\begin{equation}
Q^-(f)(t,x,v)=a^2\int_{{\bf R}^3\times S_+^2}f(t,x,v)f(t,x+a\omega,w)B(v-w, \omega) d\omega dw
\label{eekm}
\end{equation}
In equations (\ref{eekp}) and (\ref{eekm})  $S_+^2=\{\omega\in S^2 : \omega(v-w)\geq 0\}$  
is a subset of a unit sphere surface $S^2$ in ${\bf R}^3,$ 
$a$ is a diameter of hard sphere ($a\geq 0$), $\omega$ is a unit vector along the line 
passing through the centers of the spheres at their interaction,   $(v^\prime,w^\prime)$ are velocities 
after collision of two particles have precollisional velocities $(v,w)$ and 
$B(v-w, \omega)=\max(0,(v-w)\omega)$ is the collision kernel.  

This equation (\ref{ee}) is indeed the commonly known Enskog-Boltzmann equation 
and it is a modification of the original work of Enskog \cite{e22}.  
There are other different versions of the Enskog equation 
in order that they formally satisfy  
some such properties as entropy bound and 
consistence with irreversible thermodynamics 
 (see \cite{a90}, \cite{p89}, \cite{s64}, \cite{ve73}). It is worth mentioning that equation 
(\ref{ee}) still obeys 
the conservation laws of mass, momentum and energy \cite{p01}. 

As for the Boltzmann equation, two colliding particles obey the conservation laws of 
both kinetic momentum and energy,  as follows. 
\begin{equation}
v+w= v^\prime +w^\prime,\hspace{1cm} v^2+w^2={v^\prime}^2 +{w^\prime}^2.
\label{vrel01}
\end{equation}
This results in their velocity relations
\begin{equation}
v^\prime=v-[(v-w)\omega]\omega,\hspace{1cm}w^\prime=w+[(v-w)\omega]\omega,
\label{vrel02}
\end{equation}
where $\omega\in S_+^2.$  This implies that $B(v^\prime-w^\prime,-\omega)=B(v-w,\omega).$ 

The Boltzmann equation models dilute gases successfully 
but it is no longer suitable for gases with high-density effects.  
 The Enskog equation, a partial differential integral equation of the hyperbolic type,  
is a model first proposed in 1922 by Enskog \cite{e22}  
as the generalization of the Boltzmann equation  
describing the dynamical behavior of the density of a moderately dense or high-density gas. 
There are many results about the global existence and uniqueness of
 the solutions to the initial value problem for the Boltzmann and the Enskog equations without external forces. 
Concerning  the Boltzmann equation in the absence of external forces,  a global solution existence  
is obtained by DiPerna \& Lions \cite{dl} for the large data but one cannot yet know whether 
the solution to the problem is unique or not; however, a global existence and uniqueness result 
is shown by  Illner \& Shinbrot \cite{is84} 
about the solutions to the initial value problem to the Boltzmann equation 
for small initial data in the infinite vacuum. 
Concerning the  Enskog  equation without external forces,  a 
global existence and uniqueness proof is given by Polewczak \cite{p89} for near-vaccum data and  
 another one shown by Arkeryd \cite{a90} for the large data. 
Some different existence results are also given by Cercignani (see \cite{c87},  \cite{c88}). 
With an analysis of the well-posedness of the initial value problem 
in unbounded domains, some global existence and uniqueness theorems is obtained by 
Toscani \& Bellomo  \cite{tb87}  about the solutions to the  Enskog equation 
in the absence of external forces for small initial data with suitable decay to 
zero at infinity in the phase space, and the asymptotic stability of the solutions 
and the influence of the external field have been discussed. For the  Enskog equation 
in the absence of external forces, 
the $L^1$ stability of solutions is first given by Cercignani \cite{c88} 
and the time-asymptotic  behaviour of solutions in the weighted $L^\infty$ is then provided 
by Polewczak (see \cite{p89}, \cite{p89sp}, \cite{p90}).  
Many other results about this subject can be found in the references in the papers mentioned above. 
 
Two so-called Lyapunov functionals recently constructed by Ha \cite{h05}  
 mathematically neither yield the time-asymptotic behaviour in the $L^1$ norm nor recover  
the $L^1$ stability for global classical solutions 
to the  Enskog equation without external forces (see Appendix \ref{appA}).
Now there are not yet any  Lyapunov functionals 
to restore the the time-asymptotic behaviour and the stability of solutions 
to the Enskog equation in the absence of external forces.
The aim of this paper is to build two Lyapunov functionals for the Enskog equation. 
One is to describe interactions between particles with various velocities 
and another is to measure the $L^1$ distance between two classical solutions. 
The former yields the time-asymptotic convergence 
of global classical solutions to the collision free motion 
while the latter is applied into the verification 
of the $L^1$ stability of global classical solutions. 

The rest of this paper is arranged as follows. 
In Section \ref{pre} some properties of the collision operator $Q$ 
of the Enskog equation (\ref{ee}) are introduced   
including both the entropy identity and the nonincreasing property of the entropy functional. 
Then in Section \ref{ab} a Lyapunov functional is constructed and its time-decay property is given together with an  
application of the time-asymptotic behaviour of any nonnegative solution to 
the Enskog equation (\ref{ee}).  
A different Lyapunov functional is defined and its time-decay property, together an application of 
the $L^1$ stability of solution to the Enskog equation (\ref{ee}),    
is provided in Section \ref{ls}. A counterexample of inequalities given by Ha is finally shown in Appendix \ref{appA}. 

\section{Preliminaries}
\label{pre}
In this section some properties of the collision operator $Q$ of the Enskog equation 
(\ref{ee}) are introduced   
including both the entropy identity and the nonincreasing property of the entropy functional, 
and the conservation law of mass is shown for the  Enskog equation. 

We consider the Enskog equation (\ref{ee}) with (\ref{eekp}) and (\ref{eekm}).
Notice that $B(v-w, \omega)=\max(0,(v-w)\omega).$ Then we have the following lemma.
\begin{lemma}
Suppose that  $Q$ is a collisional operator as defined by  (\ref{eekp}) and (\ref{eekm}).    
Let $\psi(x,v)$ be a measurable function on ${\bf R}^3\times{\bf R}^3.$ 
If $f\in C_0({\bf R}_+\times{\bf R}^3\times{\bf R}^3),$ then 
\begin{equation}
\int_{{\bf R}^3\times{\bf R}^3}\psi(x,v)Q(f)dvdx
=\frac{a^2}{2}\int_{{\bf R}^3\times{\bf R}^3\times{\bf R}^3\times S_+^2}
B(v-w,\omega)f(t,x,v)f(t,x+a\omega,w) \nonumber
\end{equation}
\begin{equation}
\times [\psi(x,v^\prime)-\psi(x,v)+\psi(x+a\omega,w^\prime)-\psi(x+a\omega,w)]
  d\omega dw dv dx
\label{identy01}
\end{equation}
and 
\begin{equation}
\int_{{\bf R}^3\times{\bf R}^3}\psi(x,v)Q(f)dvdx
=\frac{a^2}{2}\int_{{\bf R}^3\times{\bf R}^3\times{\bf R}^3\times S_+^2}
B(v-w,\omega) f(t,x-a\omega,w^\prime)f(t,x,v^\prime)
\nonumber
\end{equation}
\begin{equation}
\times [\psi(x-a\omega,w)-\psi(x-a\omega,w^\prime)+\psi(x,v)-\psi(x,v^\prime)]
d\omega dw dv dx.
\label{identy02}
\end{equation}
\label{idk}
\end{lemma}

\begin{proof}
Denote $I_g$ and $I_l$ by 
\begin{equation}
I_g=a^2\int_{{\bf R}^3\times{\bf R}^3\times{\bf R}^3\times S_+^2}
B(v-w,\omega)\psi(x,v)f(t,x,v^\prime)f(t,x-a\omega,w^\prime)
d\omega dwdvdx
\label{ig}
\end{equation}
and 
\begin{equation}
I_l=a^2\int_{{\bf R}^3\times{\bf R}^3\times{\bf R}^3\times S_+^2}
B(v-w,\omega)\psi(x,v)f(t,x,v)f(t,x+a\omega,w)
d\omega dwdvdx,
\label{il}
\end{equation}
respectively. By (\ref{eekp}) and (\ref{eekm}), we have 
\begin{equation}
\int_{{\bf R}^3\times{\bf R}^3}\psi(x,v)Q(f)dvdx=
Ig-I_l. 
\label{iq}
\end{equation}
Let us first consider  the loss integral $I_l.$ 
By exchanging $v$ and $w$ and replacing $\omega$ with $-\omega$ 
in the integral on the right side of (\ref{il}), we can get  
\begin{equation}
I_l=a^2\int_{{\bf R}^3\times{\bf R}^3\times{\bf R}^3\times S_+^2}
B(v-w,\omega)\psi(x,w)
f(t,x,w)f(t,x-a\omega,v)d\omega dwdvdx.
\label{ilc01}
\end{equation}
Replacing $x$ with $x+a\omega$ in the integral on the right side of (\ref{ilc01}) gives 
\begin{equation}
I_l=a^2\int_{{\bf R}^3\times{\bf R}^3\times{\bf R}^3\times S_+^2}
B(v-w,\omega)\psi(x+a\omega,w)f(t,x,v)f(t,x+a\omega,w)d\omega dwdvdx.
\label{ilc02}
\end{equation}
Combining (\ref{il}) and (\ref{ilc02}), we have 
 \begin{equation}
I_l=\frac{a^2}{2}\int_{{\bf R}^3\times{\bf R}^3\times{\bf R}^3\times S_+^2}
B(v-w,\omega)[\psi(x,v)+\psi(x+a\omega,w)]f(t,x,v)f(t,x+a\omega,w)d\omega dwdvdx.
\label{ilc03}
\end{equation}
Then the gain integral $I_g$ is below considered. By using the properties that 
$B(v^\prime-w^\prime,-\omega)=B(v-w,\omega)$ and that $dwdv=dw^\prime dv^\prime,$  (\ref{ig}) can be rewritten as  
\begin{equation}
I_g=a^2\int_{{\bf R}^3\times{\bf R}^3\times{\bf R}^3\times S_+^2}
B(v^\prime-w^\prime,-\omega)\psi(x,v)f(t,x,v^\prime)f(t,x-a\omega,w^\prime)d\omega dw^\prime dv^\prime dx.
\label{igc01}
\end{equation}
By exchanging $v$ and $v^\prime$, $w$ and $w^\prime,$ and replacing $\omega$ with $-\omega$ in 
the integral on the right side of (\ref{igc01}), we have 
\begin{equation}
I_g=a^2\int_{{\bf R}^3\times{\bf R}^3\times{\bf R}^3\times S_+^2}
B(v-w,\omega)\psi(x,v^\prime)f(t,x,v)f(t,x+a\omega,w)d\omega dw dv dx.
\label{igc02}
\end{equation}
By repeating the analysis of the loss integral $I_l$ discussed above,  
(\ref{igc02}) can be rechanged as 
\begin{equation}
I_g=a^2\int_{{\bf R}^3\times{\bf R}^3\times{\bf R}^3\times S_+^2}
B(v-w,\omega)\psi(x+a\omega,w^\prime)f(t,x,v)f(t,x+a\omega,w)d\omega dw dv dx.
\label{igc03}
\end{equation}
Combining (\ref{igc02}) and (\ref{igc03}), we can get  
 \begin{equation}
I_g=\frac{a^2}{2}\int_{{\bf R}^3\times{\bf R}^3\times{\bf R}^3\times S_+^2}
B(v-w,\omega)[\psi(x,v^\prime)+\psi(x+a\omega,w^\prime)]f(t,x,v)f(t,x+a\omega,w)d\omega dwdvdx.
\label{igc04}
\end{equation}
Inserting (\ref{ilc03}) and (\ref{igc04}) 
into (\ref{iq}) gives (\ref{identy01}).  

We below prove (\ref{identy02}).  By replacing $x$ with $x-a\omega$ 
in the integral on the right side of (\ref{identy01}), we know that 
\begin{equation}
\int_{{\bf R}^3\times{\bf R}^3}\psi(x,v)Q(f)dvdx
=\frac{a^2}{2}\int_{{\bf R}^3\times{\bf R}^3\times{\bf R}^3\times S_+^2}
B(v-w,\omega)f(t,x-a\omega,v)f(t,x,w)
\nonumber
\end{equation}
\begin{equation}
\times [\psi(x-a\omega,v^\prime)-\psi(x-a\omega,v)+\psi(x,w^\prime)-\psi(x,w)]d\omega dw dv dx.
\label{qpc01}
\end{equation}
Exchanging $v$ and $w$ and replacing $\omega$ with $-\omega$ 
in the integral on the right side of (\ref{qpc01}), 
we can obtain 
\begin{equation}
\int_{{\bf R}^3\times{\bf R}^3}\psi(x,v)Q(f)dvdx
=\frac{a^2}{2}\int_{{\bf R}^3\times{\bf R}^3\times{\bf R}^3\times S_+^2}
B(v-w,\omega)f(t,x+a\omega,w)f(t,x,v)
\nonumber
\end{equation}
\begin{equation}
\times [\psi(x+a\omega,w^\prime)-\psi(x+a\omega,w)+\psi(x,v^\prime)-\psi(x,v)]d\omega dw dv dx.
\label{qpc02}
\end{equation}
Notice that $B(v-w,\omega)=B(v^\prime-w^\prime,-\omega)$ and that $dwdv=dw^\prime dv^\prime.$ 
Then (\ref{qpc02}) gives  
\begin{equation}
\int_{{\bf R}^3\times{\bf R}^3}\psi(x,v)Q(f)dvdx
=\frac{a^2}{2}\int_{{\bf R}^3\times{\bf R}^3\times{\bf R}^3\times S_+^2}
B(v^\prime-w^\prime,-\omega)f(t,x+a\omega,w)f(t,x,v)
\nonumber
\end{equation}
\begin{equation}
\times [\psi(x+a\omega,w^\prime)-\psi(x+a\omega,w)+\psi(x,v^\prime)-\psi(x,v)]d\omega dw^\prime dv^\prime dx.
\label{qpc03}
\end{equation}
By exchanging $v$ and $v^\prime$, $w$ and $w^\prime,$ and replacing $\omega$ with $-\omega$ in 
the integral on the right side of (\ref{qpc03}), 
 (\ref{identy02}) follows. 
\end{proof}

We can also know that if $\psi(x,v)=1,~v,~v^2/2$ in Lemma \ref{idk} then 
\begin{equation}
\int_{{\bf R}^3\times{\bf R}^3}\psi(x,v)Q(f)dvdx=0.
\nonumber
\end{equation}
This formally implies 
the conservation laws of mass, momentum and energy hold 
for the Enskog equation in the absence of external forces,  i.e., 
\begin{equation}
\frac{d}{dt}\int_{{\bf R}^3\times{\bf R}^3}\left( 1,v,\frac{v^2}{2}\right)f(t,x,v)dvdx=0.
\nonumber
\end{equation}

We now introduce a notation $f^\#(t,x,v)=f(t,x+vt,v).$ 
Then the Enskog equation (\ref{ee}) can be also written as  
\begin{equation}
\frac{\partial f^\#(t,x,v)}{\partial t}=Q(f)^\#(t,x,v).
\label{ee01}
\end{equation}
Then let us consider two functionals $H_B(t)$ and $I(t)$ defined as follows:
\begin{equation}
H_B(t)=\int_{{\bf R}^3\times{\bf R}^3}f(t,x,v)\ln f(t,x,v) dx dv
\label{enb}
\end{equation}
and 
\begin{equation}
I(t)=\frac{a^2}{2}\int_{{\bf R}^3\times{\bf R}^3\times{\bf R}^3\times S_+^2}
B(v-w,\omega)[f(t,x,v)f(t,x+a\omega,w)-f(t,x,v)f(t,x-a\omega,w)]d\omega dwdx dv.
\label{eni}
\end{equation}
They  have the following property.
\begin{lemma}[\cite{p89sp}] 
Assume that $H(t)=H_B(t)+\int_0^tI(s)ds $ where $H_B(t)$ and $I(t)$ are given 
by (\ref{enb}) and (\ref{eni}). If $f=f(t,x,v)$ is a nonnegative classical solution 
to the Enskog equation (\ref{ee}) with (\ref{eekp}) and (\ref{eekm}), 
then $\frac{dH(t)}{dt}\leq 0.$ 
\label{en}
\end{lemma}
\begin{proof} 
By (\ref{enb}), we have 
\begin{equation}
\frac{dH_B(t)}{dt}=\int_{{\bf R}^3\times{\bf R}^3}[1+\ln f^\#(t,x,v)]\frac{\partial f^\#(t,x,v)}{\partial t} dx dv.
\label{enbp01}
\end{equation}
Inserting (\ref{ee01}) into (\ref{enbp01}) gives 
\begin{equation}
\frac{dH_B(t)}{dt}=\int_{{\bf R}^3\times{\bf R}^3}[1+\ln f^\#(t,x,v)]Q(f)^\#(t,x,v)dx dv.
\label{enbp02}
\end{equation}
By integrating over the variable $x$ on the right side of (\ref{enbp02}), it follows that 
\begin{equation}
\frac{dH_B(t)}{dt}=\int_{{\bf R}^3\times{\bf R}^3}[1+\ln f(t,x,v)]Q(f)(t,x,v)dx dv.
\label{enbp03}
\end{equation}
By Lemma \ref{idk}, 
(\ref{enbp03}) can be rechanged as 
\begin{equation}
\frac{dH_B(t)}{dt}=\frac{a^2}{2}\int_{{\bf R}^3\times{\bf R}^3\times{\bf R}^3\times S_+^2}
B(v-w,\omega)f(t,x,v)f(t,x+a\omega,w) 
\nonumber
\end{equation}
\begin{equation}
\times \ln \left[\frac{f(t,x,v^\prime)f(t,x+a\omega,w^\prime)}{f(t,x,v)f(t,x+a\omega,w)}\right]d\omega dwdx dv.
\label{enbp04}
\end{equation}
Equation (\ref{enbp04}) is also called entropy indentity. 
Since $\ln x\leq x-1$ for $x>0,$ 
the estimation of the integral on the right side of the entropy indentity (\ref{enbp04}) reads 
\begin{equation}
\frac{dH_B(t)}{dt}\leq \frac{a^2}{2}\int_{{\bf R}^3\times{\bf R}^3\times{\bf R}^3\times S_+^2}
B(v-w,\omega)
\nonumber 
\end{equation}
\begin{equation}
\times[f(t,x,v^\prime)f(t,x+a\omega,w^\prime)
-f(t,x,v)f(t,x+a\omega,w)]d\omega dwdx dv,
\label{enbp05}
\end{equation}
i.e., $\frac{dH_B(t)}{dt}\leq -I(t).$ Therefore $\frac{dH(t)}{dt}\leq -I(t)+I(t)=0.$ 
The proof of this lemma is finished. 
\end{proof}

\section{Asymptotic Behaviour} 
\label{ab}
In this section two new functionals are built. One of them is a Lyapunov functional 
and its time-decay property is related to another functional. 
This time-decay property can be applied into our description of  
the time-asymptotic behaviour of any nonnegative solution 
to the Enskog equation (\ref{ee}) with (\ref{eekp}) and (\ref{eekm}). 

Let us begin with two new functionals $\mathfrak{D}=\mathfrak{D}^++\mathfrak{D}^-$  
and  $\mathfrak{F}=\mathfrak{F}^+ +\mathfrak{F}^-,$ 
where $\mathfrak{D}^\pm$ and $\mathfrak{F}^\pm$ are defined 
as follows:
\begin{equation}
\mathfrak{D}^\pm[f](t)=\int_{{\bf R}^3\times{\bf R}^3\times{\bf R}^3}f^\#(t,x,v)
\nonumber
\end{equation}
\begin{equation}
\times \left[\int_{{\bf R}_+\times S_+^2}
f^\#(t,x+(v-w)t\mp a\omega+\tau\frac{v-w}{(v-w)\omega},w)
d\tau d\omega \right]dwdvdx,
\label{dfun}
\end{equation}
\begin{equation}
\mathfrak{F}^\pm[f](t)=\int_{{\bf R}^3\times{\bf R}^3\times{\bf R}^3\times S_+^2}
|(v-w)\omega|f^\#(t,x,v)f^\#(t,x+(v-w)t\mp a\omega,w)d\omega dwdvdx.
\label{ffun}
\end{equation}
Obviously, by (\ref{eekp}) and (\ref{eekm}), we know that 
$\int_{{\bf R}^3\times{\bf R}^3}Q^\pm(f)^\#(t,x,v)dvdx\leq a^2\mathfrak{F}^\pm[f](t).$ 
For any nonnegative solution $f$ to the Enskog equation, 
$\mathfrak{D}[f](t)$ describes interactions between particles with various velocities. 
It can be below found that the time-decay property of $\mathfrak{D}[f](t)$ 
for any nonnegative solution $f$ to the Enskog equation 
leads directly to the time-asymptotic behaviour of this solution in $L^1({\bf R}^3\times{\bf R}^3)$ 
and so $\mathfrak{D}$ is called Lyapunov functional.  
To estimate the time decay of $\mathfrak{D}[f](t),$ we first have to show the following lemma.
\begin{lemma}
Let $F(x)$ be an integrable function on ${\bf R}^3$ 
and $v$ a vector in ${\bf R}^3.$ Assume that $S_+^2=\{\omega | v\omega\geq 0, \omega \in S^2\}$  
where $S^2$ is a unit sphere surface in ${\bf R}^3.$ Then 
\begin{equation}
\int_{{\bf R}^3}F(x)dx=a^2\int_{{\bf R}\times S_+^2}F(a\omega+\tau \frac{v}{v\omega})d\tau d\omega
\label{intfun}
\end{equation}
where $a\not=0.$ 
\label{inttrans}
\end{lemma}
\begin{proof} 
Denote $x$ by $x=(x_1,x_2,x_3).$ 
Using a transformation $x=a\omega+\tau v/(v\omega)$ where 
$\omega=(\sin\theta\cos\varphi, \sin\theta\sin\varphi,\cos\theta),$  
$\theta$ is an angle between $v$ and $\omega,$ $0\leq \theta\leq\frac{\pi}{2},$ 
$0\leq\varphi\leq2\pi.$ 
we can know that the Jacobian determinant of this transformation is 
$\frac{\partial(x_1,x_2,x_3)}{\partial(\tau,\theta,\varphi)}=a^2\sin\theta.$ 
 Lemma \ref{inttrans} therefore holds.
\end{proof}
Combining (\ref{dfun}) and Lemma \ref{inttrans}, we can easily deduce that 
$\mathfrak{D}^\pm[f](t)\leq \frac{1}{a^2}||f||_{L^1}^2$ 
for any nonnegative integrable function $f=f(t,x,v).$ 
Furthermore, we also obtain the following time-decay property of $\mathfrak{D}[f](t).$
\begin{theorem} 
Let $\mathfrak{D}$ and $\mathfrak{F}$ be defined by (\ref{dfun}) and (\ref{ffun}).  
 Assume that  $f=f(t,x,v)$ is a nonnegative classical solution 
to the Enskog equation (\ref{ee}) with (\ref{eekp}) and (\ref{eekm}) through an initial datum $f_0=f_0(x,v)$ 
and that $f\in L^1({\bf R}^3\times{\bf R}^3).$ then
\begin{equation}
\frac{d\mathfrak{D}[f](t)}{dt}+(1-4||f_0||_{L^1})\mathfrak{F}[f](t)\leq 0
\label{intesti}
\end{equation}
for any $t\in {\bf R}_+.$ 
\label{esti}
\end{theorem}
\begin{proof}
We first calculate $\frac{d\mathfrak{D}^+[f](t)}{dt}.$ To do this, we have to deduce 
that 
\begin{equation}
\partial_t\left[f^\#(t,x,v)f^\#(t,x+(v-w)t- a\omega+\tau\frac{v-w}{(v-w)\omega},w)\right]
\nonumber 
\end{equation}
\begin{equation}
=\partial_\tau\left[(v-w)\omega f^\#(t,x,v)f^\#(t,x+(v-w)t- a\omega+\tau\frac{v-w}{(v-w)\omega},w)\right]
\nonumber 
\end{equation}
\begin{equation}
+Q(f)^\#(t,x,v)f^\#(t,x+(v-w)t- a\omega+\tau\frac{v-w}{(v-w)\omega},w)
\nonumber 
\end{equation}
\begin{equation}
+f^\#(t,x,v)Q(f)^\#(t,x+(v-w)t- a\omega+\tau\frac{v-w}{(v-w)\omega},w)
\label{dtdp}
\end{equation}for any fixed variables $(x,v,w,\omega).$ 
In fact, (\ref{dtdp}) results from (\ref{ee01}) and the  two identities as follows: 
\begin{equation}
\partial_t \left[f^\#(t,x+(v-w)t- a\omega+\tau\frac{v-w}{(v-w)\omega},w)\right]
\nonumber 
\end{equation}
\begin{equation}
=(v-w)\partial_xf^\#(t,x+(v-w)t- a\omega+\tau\frac{v-w}{(v-w)\omega},w)
\nonumber 
\end{equation}
\begin{equation}
+Q(f)^\#(t,x+(v-w)t- a\omega+\tau\frac{v-w}{(v-w)\omega},w)
\label{ee02}
\end{equation} 
and
\begin{equation}
(v-w)\partial_xf^\#(t,x+(v-w)t- a\omega+\tau\frac{v-w}{(v-w)\omega},w)
\nonumber 
\end{equation}
\begin{equation}
=(v-w)\omega\partial_\tau
\left[ f^\#(t,x+(v-w)t- a\omega+\tau\frac{v-w}{(v-w)\omega},w)\right]. 
\label{dxtaueq}
\end{equation} 
Denote two functionals $\mathfrak{I}_i^\pm$ $(i=1,2)$ by 
\begin{equation}
\mathfrak{I}_1^\pm[f](t)=\int_{{\bf R}^3\times{\bf R}^3\times{\bf R}^3\times S_+^2 \times{\bf R}_+}Q(f)^\#(t,x,v)
\nonumber 
\end{equation}
\begin{equation}
\times f^\#(t,x+(v-w)t\mp a\omega+\tau\frac{v-w}{(v-w)\omega},w)d\tau d\omega dwdvdx
\label{i1fun}
\end{equation}
and 
\begin{equation}
\mathfrak{I}_2^\pm[f](t)=\int_{{\bf R}^3\times{\bf R}^3\times{\bf R}^3\times S_+^2 \times{\bf R}_+}f^\#(t,x,v)
\nonumber 
\end{equation}
\begin{equation}
\times Q(f)^\#(t,x+(v-w)t\mp a\omega+\tau\frac{v-w}{(v-w)\omega},w)d\tau d\omega dwdvdx.
\label{i2fun}
\end{equation}
By integrating (\ref{dtdp}) with respect to variables $(\tau,\omega,w,v,x),$ 
we can obtain the following identity
\begin{equation}
\frac{d\mathfrak{D}^+[f](t)}{dt}=-\mathfrak{F}^+[f](t)+\mathfrak{I}_1^+[f](t)+\mathfrak{I}_2^+[f](t)
\label{dtdp01}
\end{equation}
where $\mathfrak{F}^+$ is a functional as given by (\ref{ffun}) and 
$\mathfrak{I}_i^+$ $(i=1,2)$ are two functionals defined by (\ref{i1fun}) and (\ref{i2fun}).
We also have a similar identity
\begin{equation}
\frac{d\mathfrak{D}^-[f](t)}{dt}=-\mathfrak{F}^-[f](t)+\mathfrak{I}_1^-[f](t)+\mathfrak{I}_2^-[f](t)
\label{dtdm01}
\end{equation}
where $\mathfrak{F}^-$ is a functional as given by (\ref{ffun}) and 
$\mathfrak{I}_i^-$ $(i=1,2)$ are two functionals defined by (\ref{i1fun}) and (\ref{i2fun}).
Estimation of (\ref{i1fun}) and (\ref{i2fun}) gives 
\begin{equation}
\mathfrak{I}_i^+[f](t)+\mathfrak{I}_i^-[f](t)
\leq 2||f_0||_{L^1}\mathfrak{F}^+[f](t)
\label{iesti}
\end{equation}
since $Q(f)^\#(t,x,v)\leq Q^+(f)^\#(t,x,v).$ 
Combining (\ref{dtdp01}), (\ref{dtdm01}) and (\ref{iesti}), we hence know that (\ref{intesti}) holds.
This completes the proof of this theorem.
\end{proof}
By Theorem \ref{esti}, we then get 
\begin{theorem} 
Put $f_\infty(x,v)=f_0(x,v)+\int_{{\bf R}_+}Q(f)^\#(s,x,v)ds.$  
Assume that  $f=f(t,x,v)$ is a nonnegative classical solution 
to the Enskog equation (\ref{ee}) with (\ref{eekp}) and (\ref{eekm}) through an initial datum $f_0=f_0(x,v)$ 
satisfying $||f_0||_{L^1}< 1/4$ and that $f\in L^1({\bf R}^3\times{\bf R}^3).$ 
Then 
$f(t,x,v)$ converges in $L^1({\bf R}^3\times{\bf R}^3)$ to $f_\infty(x-vt,v)$ as $t\to 0.$ 
\label{ths}
\end{theorem}
\begin{proof}
By Theorem \ref{esti}, we know that  
$ \frac{d\mathfrak{D}[f](t)}{dt}
+(1-4||f_0||_{L^1})\mathfrak{F}[f](t)\leq 0 $ 
where $\mathfrak{D}[f](t)$ and $\mathfrak{F}[f](t)$ are the same as in Lemma \ref{esti}. 
It follows that  
\begin{equation}
\mathfrak{D}[f](t) 
+(1-4||f_0||_{L^1})\int_0^t\mathfrak{F}[f](s)ds
\leq \mathfrak{D}[f](0) 
\nonumber
\end{equation}
for any $t\in {\bf R}_+,$ which implies that 
$
\int_{{\bf R}^3\times{\bf R}^3\times{\bf R}_+}
Q^\pm(f)(t,x,v)dxdvdt<+\infty.
$ 
Hence it can be easily found that 
\begin{equation}
\int_{{\bf R}^3\times{\bf R}^3}
|f^\#(t,x,v)-f_{\infty}(x,v)|dxdv\leq
\int_t^{+\infty}\int_{{\bf R}^3\times{\bf R}^3}
[Q^+(f)(s,x,v)+Q^-(f)(s,x,v)]dxdvds\to 0
\nonumber
\end{equation}
as $t\to +\infty.$
Thus $f(t,x,v)\to f_{\infty}(x-vt,v)$ in $L^1$ as $t\to +\infty.$
This completes our proof of Theorem \ref{ths}.
\end{proof}
In Theorem \ref{ths} the time-asymptotic 
behaviour of any solution to the Enskog equation 
is in fact the time-asymptotic convergence of this solution in the $L^1$ norm to 
the free motion as $t$ trends to infinity.  
The time-asymptotic convergence in the $L^\infty$ norm is shown by Polewczak \cite{p89}. 

\section{$L^1$ Stability}
\label{ls}
In this section some new functionals are constructed 
for the $L^1$ stability of  
 global classical solutions to the Enskog equation (\ref{ee}) 
with (\ref{eekp}) and (\ref{eekm}).  
One of them is a Lyapunov functional 
and it is equivalent to the $L^1$ distance functional. 
The time-decay property of the Lyapunov functional 
is also shown for the $L^1$ stability. 

Let us begin by constructing two functionals  
$\mathfrak{L}[f,g](t)$ and $\mathfrak{F}_d[f,g](t)$ as follows. 
$\mathfrak{L}[f,g](t)$ is denoted by 
\begin{equation}
\mathfrak{L}[f,g](t)=\left\{1+k_1(\mathfrak{D}[f](t)+\mathfrak{D}[g](t))\right\}
\mathfrak{L}_d[f,g](t)+k_2\mathfrak{D}_d[f,g](t)  
\label{lfun}
\end{equation}
where $k_1$ and $k_2$ are positive constants to be determined later, 
$\mathfrak{D}$ is the same as given by (\ref{dfun}),  $\mathfrak{L}_d$ is denoted by 
\begin{equation}
\mathfrak{L}_d[f,g](t)=\int_{{\bf R}^3\times{\bf R}^3}|f-g|^\#(t,x,v)dxdv
\label{ldfun}
\end{equation}
and $\mathfrak{D}_d[f,g](t)=\mathfrak{D}_d^+[f,g](t)+\mathfrak{D}_d^-[f,g](t)$ 
with 
\begin{equation}
\mathfrak{D}_d^\pm[f,g](t)=\int_{{\bf R}^3\times{\bf R}^3}|f-g|^\#(t,x,v)
\nonumber
\end{equation}
\begin{equation}
\times \left[\int_{{\bf R}^3\times{\bf R}_+\times S_+^2}
(f+g)^\#(t,x+t(v-w)\mp a\omega+\tau\frac{v-w}{(v-w)\omega},w)d\tau d\omega dw\right]dxdv. 
\label{ddfun}
\end{equation}
$\mathfrak{F}_d[f,g](t)$ is defined by 
$\mathfrak{F}_d[f,g](t)=\mathfrak{F}_d^+[f,g](t)+\mathfrak{F}_d^-[f,g](t)$ with 
\begin{equation}
\mathfrak{F}_d^\pm[f,g](t)=\int_{{\bf R}^3\times{\bf R}^3\times{\bf R}^3\times S_+^2}(v-w)\omega|f-g|^\#(t,x,v)
\nonumber
\end{equation}
\begin{equation}
\times (f+g)^\#(t,x+t(v-w)\mp a\omega,w) d\omega dwdxdv. 
\label{fdfun}
\end{equation} 
This functional $\mathfrak{L}$ is here called Lyapunov functional. 
Then we have the following property of the equivalence between the Lyapunov and 
$L^1$ distance functionals $\mathfrak{L}$ and $\mathfrak{L}_d.$
\begin{lemma}
Let $\mathfrak{L}$ and $\mathfrak{L}_d$ be defined by (\ref{lfun}) and (\ref{ldfun}) respectively.  
 Assume that  $f=f(t,x,v)$  and $g=g(t,x,v)$ are two nonnegative classical solutions  
to the Enskog equation (\ref{ee}) with (\ref{eekp}) and (\ref{eekm}) through initial data $f_0=f_0(x,v)$ 
and $g_0=g_0(x,v)$ respectively, and that $f$ and $g$ 
are elements in $L^1({\bf R}^3\times{\bf R}^3).$  Then 
\begin{equation}
\mathfrak{L}_d[f,g](t)\leq \mathfrak{L}[f,g](t)\leq C_0\mathfrak{L}_d[f,g](t)
\label{equivineq}
\end{equation}
for any $t\in {\bf R}_+,$  where 
$C_0=1+2k_1(||f_0||_{L^1}^2+||g_0||_{L^1}^2)/a^2+2k_2(||f_0||_{L^1}+||g_0||_{L^1}).$ 
\label{equiv}
\end{lemma}
\begin{proof}
The definition (\ref{lfun}) of $\mathfrak{L}$ can be rewritten as 
\begin{equation}
\mathfrak{L}[f,g](t)=\int_{{\bf R}^3\times{\bf R}^3}|f-g|^\#(t,x,v)W^\#(t,x,v)dxdv
\nonumber
\end{equation} 
where $W(t,x,v)$ is defined by
\begin{equation}
W(t,x,v)=1+k_1[\mathfrak{D}[f](t)+\mathfrak{D}[g](t)]
\nonumber
\end{equation}
\begin{equation}
+k_2\int_{{\bf R}^3\times{\bf R}_+\times S_+^2} 
(f+g)(t,x-a\omega+\tau\frac{v-w}{(v-w)\omega},w)d\tau d\omega dw
\nonumber
\end{equation}
\begin{equation}
+k_2\int_{{\bf R}^3\times{\bf R}_+\times S_+^2} 
(f+g)(t,x+a\omega+\tau\frac{v-w}{(v-w)\omega},w)d\tau d\omega dw,
\nonumber
\end{equation}
and by Lemma \ref{inttrans} it can be found that 
$W(t,x,v)$ is bounded by the $L^1$ norms of $f_0$ and $g_0$ as follows:
\begin{equation}
1\leq W(t,x,v)\leq 1+2k_1(||f_0||_{L^1}^2+||g_0||_{L^1}^2)/a^2+2k_2(||f_0||_{L^1}+||g_0||_{L^1}).
\nonumber
\end{equation} 
Our proof of this lemma hence ends up. 
\end{proof}

For any solutions $f=f(t,x,v)$  and $g=g(t,x,v)$ 
to the Enskog equation (\ref{ee}) with (\ref{eekp}) and (\ref{eekm}), 
the time-decay properties of the two functionals $\mathfrak{D}_d[f,g](t)$ 
and $\mathfrak{L}_d[f,g](t)$ can be also obtained as follows.
\begin{lemma} 
Let $\mathfrak{L}$ and $\mathfrak{L}_d$ be defined by (\ref{lfun}) and (\ref{ldfun}) respectively.  
 Assume that  $f=f(t,x,v)$  and $g=g(t,x,v)$ are two nonnegative classical solutions  
to the Enskog equation (\ref{ee}) with (\ref{eekp}) and (\ref{eekm}) through initial data $f_0=f_0(x,v)$ 
and $g_0=g_0(x,v)$ respectively, and that $f$ and $g$ 
are elements in $L^1({\bf R}^3\times{\bf R}^3).$ Then 
\begin{equation}
\frac{d\mathfrak{L}_d[f,g](t)}{dt}\leq a^2\mathfrak{F}_d[f,g](t)
\label{ldesti}
\end{equation}
and 
\begin{equation}
\frac{d\mathfrak{D}_d[f,g](t)}{dt}+[1-2(||f_0||_{L^1}+||g_0||_{L^1})]
\mathfrak{F}_d[f,g](t)\leq 2\{\mathfrak{F}[f](t)+\mathfrak{F}[g](t)\}\mathfrak{L}_d[f,g](t)
\label{ddesti}
\end{equation}
for any $t\in {\bf R}_+,$ where $\mathfrak{F}$ and $\mathfrak{F}_d$ are given by (\ref{ffun}) and (\ref{fdfun}) respectively. 
\label{ldddesti}
\end{lemma}
\begin{proof}
Let us first prove (\ref{ldesti}). By (\ref{ee01}), we have the following inequalities 
\begin{equation}
\frac{\partial [f^\#(t,x,v)-g^\#(t,x,v)]}{\partial t}\leq\mathfrak{R}[f,g]^\#(t,x,v)
\label{eeineq01}
\end{equation}
and
\begin{equation}
\frac{\partial [g^\#(t,x,v)-f^\#(t,x,v)]}{\partial t}\leq\mathfrak{R}[f,g]^\#(t,x,v)
\label{eeineq02}
\end{equation}
for any nonnegative solutions $f$ and $g$ to the Enskog equation, 
 where $\mathfrak{R}$ is denoted by 
\begin{equation}
\mathfrak{R}[f,g](t,x,v)=\frac{a^2}{2}\int_{{\bf R}^3\times S_+^2}
(v-w)\omega[|f-g|(t,x,v^\prime)(f+g)(t,x-a\omega,w^\prime)
\nonumber
\end{equation}
\begin{equation}
+ |f-g|(t,x-a\omega,w^\prime)(f+g)(t,x,v^\prime)
\nonumber
\end{equation}
\begin{equation}
+ |f-g|(t,x,v)(f+g)(t,x+a\omega,w)+|f-g|(t,x+a\omega,w)(f+g)(t,x,v)]d\omega dw. 
\label{eeineqk}
\end{equation}
Using (\ref{eeineq01}) and (\ref{eeineq02}), we have 
\begin{equation}
\frac{d\mathfrak{L}_d[f,g](t)}{dt}\leq\int_{{\bf R}^3\times{\bf R}^3}\mathfrak{R}[f,g]^\#(t,x,v)dxdv=a^2\mathfrak{F}_d[f,g](t).
\nonumber
\end{equation}
Hence (\ref{ldesti}) holds.  Then we below prove (\ref{ddesti}). 
To do this, we have to calculate $\frac{d\mathfrak{D}_d^\pm[f,g](t)}{dt}.$

We first calculate $\frac{d\mathfrak{D}_d^+[f,g](t)}{dt}.$ To do this, we have to deduce 
that 
\begin{equation}
\partial_t\left[(f-g)^\#(t,x,v)(f+g)^\#(t,x+(v-w)t- a\omega+\tau\frac{v-w}{(v-w)\omega},w)\right]
\nonumber 
\end{equation}
\begin{equation}
\leq\partial_\tau\left[(v-w)\omega (f-g)^\#(t,x,v)(f+g)^\#(t,x+(v-w)t- a\omega+\tau\frac{v-w}{(v-w)\omega},w)\right]
\nonumber 
\end{equation}
\begin{equation}
+\mathfrak{R}[f,g]^\#(t,x,v)(f+g)^\#(t,x+(v-w)t- a\omega+\tau\frac{v-w}{(v-w)\omega},w)
\nonumber 
\end{equation}
\begin{equation}
+(f-g)^\#(t,x,v)[Q(f)+Q(g)]^\#(t,x+(v-w)t- a\omega+\tau\frac{v-w}{(v-w)\omega},w)
\label{dtddp01}
\end{equation}
and 
\begin{equation}
\partial_t\left[(g-f)^\#(t,x,v)(f+g)^\#(t,x+(v-w)t- a\omega+\tau\frac{v-w}{(v-w)\omega},w)\right]
\nonumber 
\end{equation}
\begin{equation}
\leq\partial_\tau\left[(v-w)\omega (g-f)^\#(t,x,v)(f+g)^\#(t,x+(v-w)t- a\omega+\tau\frac{v-w}{(v-w)\omega},w)\right]
\nonumber 
\end{equation}
\begin{equation}
+\mathfrak{R}[f,g]^\#(t,x,v)(f+g)^\#(t,x+(v-w)t- a\omega+\tau\frac{v-w}{(v-w)\omega},w)
\nonumber 
\end{equation}
\begin{equation}
+(g-f)^\#(t,x,v)[Q(f)+Q(g)]^\#(t,x+(v-w)t- a\omega+\tau\frac{v-w}{(v-w)\omega},w)
\label{dtddp02}
\end{equation}
for any fixed variables $(x,v,w,\omega).$ 
In fact, (\ref{dtddp01}) and (\ref{dtddp02}) result from (\ref{ee01}), (\ref{ee02}), (\ref{dxtaueq}), (\ref{eeineq01}),  
 (\ref{eeineq02}) and the  two identities as follows: 
\begin{equation}
\partial_t \left[g^\#(t,x+(v-w)t- a\omega+\tau\frac{v-w}{(v-w)\omega},w)\right]
\nonumber 
\end{equation}
\begin{equation}
=(v-w)\partial_xg^\#(t,x+(v-w)t- a\omega+\tau\frac{v-w}{(v-w)\omega},w)
\nonumber 
\end{equation}
\begin{equation}
+Q(g)^\#(t,x+(v-w)t- a\omega+\tau\frac{v-w}{(v-w)\omega},w)
\nonumber
\end{equation} 
and
\begin{equation}
(v-w)\partial_xg^\#(t,x+(v-w)t- a\omega+\tau\frac{v-w}{(v-w)\omega},w)
\nonumber 
\end{equation}
\begin{equation}
=(v-w)\omega\partial_\tau
\left[ g^\#(t,x+(v-w)t- a\omega+\tau\frac{v-w}{(v-w)\omega},w)\right]. 
\nonumber
\end{equation}
Denote $\mathfrak{J}_i^\pm[f,g](t)$ ($i=1,2$) by 
\begin{equation}
\mathfrak{J}_1^\pm[f,g](t)=
\int_{{\bf R}^3\times{\bf R}^3\times{\bf R}^3\times S_+^2 \times{\bf R}_+}
\mathfrak{R}[f,g]^\#(t,x,v)
\nonumber 
\end{equation}
\begin{equation}
\times (f+g)^\#(t,x+(v-w)t\mp a\omega+\tau\frac{v-w}{(v-w)\omega},w)
d\tau d\omega dwdvdx 
\label{rfgfun}
\end{equation}
and 
\begin{equation}
\mathfrak{J}_2^\pm[f,g](t)=
\int_{{\bf R}^3\times{\bf R}^3\times{\bf R}^3\times S_+^2 \times{\bf R}_+}
|f-g|^\#(t,x,v)
\nonumber 
\end{equation}
\begin{equation}
\times (Q(f)+Q(g))^\#(t,x+(v-w)t\mp a\omega+\tau\frac{v-w}{(v-w)\omega},w)
d\tau d\omega dwdvdx. 
\label{qfgfun}
\end{equation}
Then, using (\ref{dtddp01}) and (\ref{dtddp02}),  
we can obtain the following inequality
\begin{equation}
\frac{d\mathfrak{D}_d^+[f,g](t)}{dt}\leq -\mathfrak{F}_d^+[f,g](t)
+\mathfrak{J}_1^+[f,g](t)+\mathfrak{J}_2^+[f,g](t).
\label{dtddpineq}
\end{equation}
Similarly calculating $\frac{d\mathfrak{D}_d^-[f,g](t)}{dt},$  we also have 
\begin{equation}
\frac{d\mathfrak{D}_d^-[f,g](t)}{dt}\leq -\mathfrak{F}_d^-[f,g](t)
+\mathfrak{J}_1^-[f,g](t)+\mathfrak{J}_2^-[f,g](t).
\label{dtddmineq}
\end{equation}
Combining (\ref{dtddpineq}) and (\ref{dtddmineq}), we get 
\begin{equation}
\frac{d\mathfrak{D}_d[f,g](t)}{dt}\leq -\mathfrak{F}_d[f,g](t)
+\mathfrak{J}_1^+[f,g](t)+\mathfrak{J}_1^-[f,g](t)+\mathfrak{J}_2^+[f,g](t)+\mathfrak{J}_2^-[f,g](t).
\label{dtddineq}
\end{equation}
By (\ref{rfgfun}) and (\ref{qfgfun}), the estimates of 
$\mathfrak{J}_i^+[f,g](t)+\mathfrak{J}_i^-[f,g](t)$ ($i=1,2$) can be obtained as follows: 
\begin{equation}
\mathfrak{J}_1^+[f,g](t)+\mathfrak{J}_1^-[f,g](t)
\leq 2(||f_0||_{L^1}+||g_0||_{L^1})\mathfrak{F}_d^+[f,g](t)
\label{jfgesti01}
\end{equation} 
and
\begin{equation}
\mathfrak{J}_2^+[f,g](t)+\mathfrak{J}_2^-[f,g](t)\leq 
2(\mathfrak{F}[f](t)+\mathfrak{F}[g](t))\mathfrak{L}_d[f,g](t).
\label{jfgesti02}
\end{equation}
By applying  (\ref{jfgesti01}) and (\ref{jfgesti02}) into  
the estimation of the right side of (\ref{dtddineq}),  (\ref{ddesti}) follows.
\end{proof}

It can be further shown that the functional $\mathfrak{L}$ has the following time-decay property.  
\begin{theorem} 
Let $\mathfrak{L}$ and $\mathfrak{F}_d$ be defined by (\ref{lfun}) and (\ref{fdfun}) respectively.  
 Assume that  $f=f(t,x,v)$  and $g=g(t,x,v)$ are two nonnegative classical solutions  
to the Enskog equation (\ref{ee}) with (\ref{eekp}) and (\ref{eekm}) through initial data $f_0=f_0(x,v)$ 
and $g_0=g_0(x,v)$ satisfying $||f_0||_{L^1}< (2-\sqrt{2})/4$ and $||g_0||_{L^1}< (2-\sqrt{2})/4$ respectively,
and that $f$ and $g$ are elements in $L^1({\bf R}^3\times{\bf R}^3).$ 
Then there exists a positive constant $\alpha \in (0, 2\sqrt{2})$ such that    
\begin{equation}
\frac{d\mathfrak{L}[f,g](t)}{dt}
\leq a^2\left[\frac{\alpha^2-8}{4(2+\alpha)}k+1\right]
\mathfrak{F}_d[f,g](t)
\label{lftd}
\end{equation}
with $k_1=(2+\alpha)k_2$ and $k_2=ka^2$ 
for any $t\in {\bf R}_+,$ where $k>\frac{4(2+\alpha)}{8-\alpha^2}.$ 
 Furthermore, the Lyapunov type estimate holds as follows: 
\begin{equation}
\mathfrak{L}[f,g](t)+C\int_0^t\mathfrak{F}_d[f,g](s)ds\leq \mathfrak{L}[f,g](0)
\label{lfdesti}
\end{equation}
for any $t\in {\bf R}_+,$ where $C$ is a positive constant independent of $t.$ 
\label{lfdestith}
\end{theorem}
\begin{proof}
By the definition (\ref{lfun}) of the functional $\mathfrak{L},$  
using Theorem \ref{esti} and Lemma \ref{ldddesti}, we have 
\begin{equation}
\frac{d\mathfrak{L}[f,g](t)}{dt}
=\left[1+k_1(\mathfrak{D}[f](t)+\mathfrak{D}[g](t))\right]\frac{d\mathfrak{L}_d[f,g](t)}{dt}
\nonumber 
\end{equation}
\begin{equation}
+k_1\left(\frac{d\mathfrak{D}[f](t)}{dt}+\frac{d\mathfrak{D}[g](t)}{dt}\right)\mathfrak{L}_d[f,g](t)
+k_2\frac{d\mathfrak{D}_d[f,g](t)}{dt}
\nonumber 
\end{equation}
\begin{equation}
\leq \left[a^2k_1(\mathfrak{D}[f](t)+\mathfrak{D}[g](t))
+2k_2(||f_0||_{L^1}+||g_0||_{L^1})-k_2+a^2\right]
\mathfrak{F}_d[f,g](t)
\nonumber 
\end{equation}
\begin{equation}
+\left[4k_1||f_0||_{L^1}-k_1+2k_2\right]
\mathfrak{F}[f](t)\mathfrak{L}_d[f,g](t)
+\left[4k_1||g_0||_{L^1}-k_1+2k_2\right]
\mathfrak{F}[g](t)\mathfrak{L}_d[f,g](t).
\nonumber
\end{equation}
Since $\mathfrak{D}[f](t)\leq \frac{2}{a^2}||f_0||_{L^1}^2$ and $\mathfrak{D}[g](t)\leq \frac{2}{a^2}||g_0||_{L^1}^2$, 
it follows that 
\begin{equation}
\frac{d\mathfrak{L}[f,g](t)}{dt}
\leq \left[2k_1(||f_0||_{L^1}^2+||g_0||_{L^1}^2)
+2k_2(||f_0||_{L^1}+||g_0||_{L^1})-k_2+a^2\right]
\mathfrak{F}_d[f,g](t)
\nonumber 
\end{equation}
\begin{equation}
+\left[4k_1||f_0||_{L^1}-k_1+2k_2\right]
\mathfrak{F}[f](t)\mathfrak{L}_d[f,g](t)
+\left[4k_1||g_0||_{L^1}-k_1+2k_2\right]
\mathfrak{F}[g](t)\mathfrak{L}_d[f,g](t).
\nonumber
\end{equation}
Notice that $||f_0||_{L^1}< (2-\sqrt{2})/4$ and $||g_0||_{L^1}< (2-\sqrt{2})/4.$  
There exists a positive constant $\alpha \in (0, 2\sqrt{2})$ such that 
$||f_0||_{L^1}\leq \frac{\alpha}{4(2+\alpha)}$ 
and $||g_0||_{L^1}\leq \frac{\alpha}{4(2+\alpha)}.$
Let us choose $k_1=(2+\alpha)k_2$ and $k_2=ka^2$ 
where $k>\frac{4(2+\alpha)}{8-\alpha^2}.$ 
By estimating the right side of the above inequality, 
it can be found that 
\begin{equation}
\frac{d\mathfrak{L}[f,g](t)}{dt}
\leq a^2\left[\frac{\alpha^2-8}{4(2+\alpha)}k+1\right]\mathfrak{F}_d[f,g](t). 
\nonumber 
\end{equation}
(\ref{lftd}) and (\ref{lfdesti}) then follows.
The proof of Theorem \ref{lfdestith} is hence finished.
\end{proof}
By Theorem \ref{lfdestith}, we further have 
\begin{theorem}
Assume that  $f=f(t,x,v)$  and $g=g(t,x,v)$ are two nonnegative classical solutions  
to the Enskog equation (\ref{ee}) with (\ref{eekp}) and (\ref{eekm}) through initial data $f_0=f_0(x,v)$ 
and $g_0=g_0(x,v)$ satisfying 
$||f_0||_{L^1}< (2-\sqrt{2})/4$ and $||g_0||_{L^1}< (2-\sqrt{2})/4$ respectively, 
and that $f$ and $g$ are elements in $L^1({\bf R}^3\times{\bf R}^3).$ 
Then  
\begin{equation}
\int_{{\bf R}^3\times{\bf R}^3}|f-g|(t,x,v)dxdv
\leq C\int_{{\bf R}^3\times{\bf R}^3}|f_0-g_0|(x,v)dxdv
\label{lstability}
\end{equation}
for any $t\in {\bf R}_+,$ where $C$ is a positive constant independent of $t.$ 
\label{tht}
\end{theorem} 
\begin{proof}
By Theorem \ref{lfdestith}, there exists a positive constant $\alpha\in (0,2\sqrt{2})$ such that  
$\mathfrak{L}[f,g](t)\leq \mathfrak{L}[f,g](0)$ with $k_1=(2+\alpha)k_2$ and $k_2=ka^2,$ 
where $k>\frac{4(2+\alpha)}{8-\alpha^2}.$ 
By using Lemma \ref{equiv}, it follows that 
\begin{equation}
\mathfrak{L}_d[f,g](t)\leq\mathfrak{L}[f,g](t)\leq \mathfrak{L}[f,g](0)\leq C_0\mathfrak{L}_d[f,g](0). 
\nonumber
\end{equation}
This hence completes the proof of Theorem \ref{tht}.
\end{proof}
Theorem \ref{tht} can be here regarded as 
an application of the Lyapunov functional $\mathfrak{L}$  into 
recovering a result of Cercignani \cite{c88} about 
the $L^1$ stability of classical solutions to the Enskog equation  
under the assumption of the suitably small initial data. 

\section*{Acknowledgement} 
The author would like to thank the referee of this paper for his/her helpful comments on this work.



\appendix
\section{A Counterexample of Inequalities Given by Ha} 
\label{appA}
Below is shown a counterexample of 
two inequalities   
as follows: 
\begin{equation}
\int_{S_+^2\times {\bf R}_+}f^\#(x+t(v-v_*)\pm a\sigma+\tau n(v,v_*),v_*,t) d\tau d\sigma
\leq \int_{{\bf R}^3}f^\#(x,v_*,t)dx
\label{ineqha}
\end{equation}
where $S_+^2=\{\sigma : (v-v_*)\sigma\geq 0, \sigma\in S^2\},$ $n(v,v_*)=(v-v_*)/|v-v_*|$ 
and $f(x,v,t)$ is a nonnegative function of the space $x\in {\bf R}^3,$ 
the velocity $v\in {\bf R}^3$ and the time $t\in {\bf R}_+.$
These inequalities are given by Ha (see (3.1) and (3.2) at Page 1004 in \cite{h05}) and applied  into 
the proof of his theorems which show that 
two so-called Lyapunov functionals have their Lyapunov type estimates that   
 yield the time-asymptotic behaviour and 
the $L^1$ stability of global classical solutions 
to the  Enskog equation without external forces.  

To give a counterexample, we first have to prove the following lemma. 
\begin{lemma}
Let $F(x)$ be an integrable function on ${\bf R}^3$ 
and $n$ a unit vector in ${\bf R}^3.$ Assume that $S_+^2=\{\omega : n\omega\geq 0, \omega \in S^2\}$  
where $S^2$ is a unit sphere surface in ${\bf R}^3.$ Then 
\begin{equation}
\int_{{\bf R}^3}F(x)dx=a^2\int_{{\bf R}\times S_+^2}F(a\omega+\tau n)n\omega d\tau d\omega
\label{intfunex}
\end{equation}
where $a\not=0.$ 
\label{inttransex}
\end{lemma}
\begin{proof} Without loss of generality, we assume that $a>0.$ 
Let $\omega=(\sin\theta\cos\varphi, \sin\theta\sin\varphi,\cos\theta),$  
where 
$\theta$ is an angle between $v$ and $\omega,$ $0\leq \theta\leq\frac{\pi}{2},$ 
$0\leq\varphi\leq2\pi.$ 
Denote $n$ and $x$ by $n=(n_1,n_2,n_3)$ and $x=(x_1,x_2,x_3)$ respectively. 
Take a transformation $x=a\omega+\tau n,$ i.e., 
\begin{equation}
x_1=\tau n_1+a\sin\theta\cos\varphi,~ 
x_2=\tau n_2+a\sin\theta\sin\varphi,~
x_3=\tau n_3+a\cos\theta. 
\label{extrans}
\end{equation}
By direct calculation, it then follows that the Jacobian determinant of this transformation is 
$\frac{\partial(x_1,x_2,x_3)}{\partial(\tau,\theta,\varphi)}=a^2n\omega\sin\theta.$ 
(\ref{intfunex}) hence holds. Our proof of this lemma is thus finished. 
\end{proof}
Lemma \ref{inttransex} not only implies that $F(x)$ 
must be integrable in ${\bf R}^3$ if $F(a\omega+\tau n)$ 
is integrable in ${\bf R}\times S_+^2,$ 
but also indicates that $F(a\omega+\tau n)$ 
might not be integrable in ${\bf R}\times S_+^2$ 
even if $F(x)$ is integrable in ${\bf R}^3$
since it can be known from (\ref{extrans}) that 
$n\omega=(xn-\tau)/a$ and $(\tau-xn)^2=(xn)^2+a^2-x^2.$ 
For example, let us choose $n=(1,0,0)$ and $\hat{F}(x)=\widetilde{F}(x)/\sqrt{|x_2^2+x_3^2-a^2|},$ 
where $\widetilde{F}(x)\in C_0^\infty(\Omega),$ 
$\widetilde{F}(x)=1$ as $x\in B_{2a},$ $0\leq\widetilde{F}(x)\leq 1$ as $x\in \Omega,$ $B_{2a}\subset \Omega,$
$\Omega$ is a compact subset in ${\bf R}^3$ and 
$B_{2a}$ represents a ball with its center in the origin and its radius $2a.$ It is easy to see that 
$\hat{F}(a\omega+\tau n)$ 
is not  integrable in ${\bf R}\times S_+^2$ 
but $\hat{F}(x)$ is integrable in ${\bf R}^3.$
This is a good counterexample of inequalities (\ref{ineqha})  
and so the two inequalities do not hold mathematically.
\end{document}